\documentclass[twoside]{article}
\usepackage{fleqn,espcrc2,epsf}

\title{The charmed and bottom meson spectrum from lattice NRQCD}

\author{Randy Lewis\address{Department of Physics, University of Regina,
        Regina, SK, Canada~~S4S 0A2}
        and
        R. M. Woloshyn\address{TRIUMF, 4004 Wesbrook Mall, Vancouver, BC,
        Canada~~V6T 2A3}}
       
\begin{document}
\pagestyle{empty}

\begin{abstract}
The mass spectrum of S and P-wave mesons containing a single heavy quark
has been computed using quenched lattice nonrelativistic QCD.
Numerical results have been obtained at first, second and third order in
the heavy quark expansion, so convergence can be discussed.
The computed spectrum of charmed and bottom mesons is compared to
existing model calculations and experimental data.
\vspace{1pc}
\end{abstract}

\maketitle

\section{MOTIVATION}

The masses of the ground state doublets of S-wave heavy-light mesons
($D^{(*)}$, $D_s^{(*)}$, $B^{(*)}$ and $B_s^{(*)}$) are well known 
experimentally,
and thus provide a testing ground for the lattice method.  In particular,
they can be used to examine the convergence properties of the heavy quark 
expansion in nonrelativistic QCD (NRQCD),\cite{ourSwaves} which is the most 
economical lattice action available for heavy quark physics.\cite{NRQCD}

The spectrum of P-wave heavy-light mesons is not yet understood, 
experimentally nor theoretically, so here the lattice can offer predictions.
Do the masses follow the ordering that is observed in the hydrogen atom?
What are the magnitudes of the mass splittings among P-waves --- tens of
MeV or hundreds?

Lattice studies of the P-wave spectrum have been performed for an
infinitely-heavy quark\cite{MicP}.  A subset of the P-wave states have also 
been observed by using a relativistic (clover) action in the charmed region,
and extrapolating to the bottom region\cite{Boyle}.
Recently, the lattice NRQCD approach has been used to work directly with 
$B^{**}$ mesons\cite{ourPwaves,AliKhan,Hein} as well as 
$D^{**}$ mesons\cite{ourPwaves,Hein}.

In the following sections, the lattice NRQCD method is briefly reviewed
and the results of numerical simulations are presented.  Implications for
the convergence of the lattice NRQCD expansion are discussed, and predictions
for the $B^{**}$ and $D^{**}$ mass spectra from various lattice simulations
are compared to theoretical models.

\section{LATTICE NRQCD METHOD}

The application of lattice NRQCD to heavy-light mesons entails an expansion
of the action in inverse powers of the heavy quark mass $M$,
\begin{equation}\label{action}
S = S_{q,G} + \int{\rm d}^4x\,Q^\dagger\left(iD_t+\frac{\Delta^{(2)}}{2M}
                             -\delta{H}\right)Q,
\end{equation}
where $S_{q,G}$ has no heavy quark fields, and\cite{NRQCD,ManB}
\begin{equation}
\delta{H} = \delta{H}^{(1)} + \delta{H}^{(2)} + \delta{H}^{(3)} + O(1/M^4),
\end{equation}
\vspace{-7mm}

\begin{eqnarray}
\delta{H}^{(1)} &=& -\frac{c_4}{u_s^4}\frac{g}{2M}\mbox{{\boldmath$\sigma$}}
                    \cdot\tilde{\bf B} + c_5\frac{a_s^2\Delta^{(4)}}{24M}, \\
\delta{H}^{(2)} &=& \frac{c_2}{u_s^2u_t^2}\frac{ig}{8M^2}(\tilde{\bf \Delta}
                    \cdot\tilde{\bf E}-\tilde{\bf E}\cdot\tilde{\bf \Delta})
                    \nonumber \\
            && -\frac{c_3}{u_s^2u_t^2}\frac{g}{8M^2}\mbox{{\boldmath$\sigma$}}
               \cdot(\tilde{\bf \Delta}\times\tilde{\bf E}-\tilde{\bf E}\times
           \tilde{\bf \Delta}) \nonumber \\
               && - c_6\frac{a_t(\Delta^{(2)})^2}{16nM^2}, \\
\delta{H}^{(3)} &=& -c_1\frac{(\Delta^{(2)})^2}{8M^3}
                    -\frac{c_7}{u_s^4}\frac{g}{8M^3}\left\{\tilde\Delta^{(2)},
                        \mbox{{\boldmath$\sigma$}}\cdot\tilde{\bf B}\right\}
                    \nonumber \\
                 && -\frac{c_9ig^2}{8M^3}\mbox{{\boldmath$\sigma$}}\cdot
                     \left(\frac{\tilde{\bf E}\times\tilde{\bf E}}{u_s^4u_t^4}
                    +\frac{\tilde{\bf B}\times\tilde{\bf B}}{u_s^8}\right)
                     \nonumber \\
             && -\frac{c_{10}g^2}{8M^3}\left(\frac{\tilde{\bf E}^2}{u_s^4u_t^4}
                +\frac{\tilde{\bf B}^2}{u_s^8}\right) \nonumber \\
             && -c_{11}\frac{a_t^2(\Delta^{(2)})^3}{192n^2M^3}.\label{dH3}
\end{eqnarray}
The chromoelectric and chromomagnetic fields are 
$E_i = F_{4i}$ and $B_i = \epsilon_{ijk}F_{ijk}/2$ respectively, and $\Delta$
denotes the lattice derivative.  The coefficients $c_i$ are usually set to
their classical values ($c_i=1$) in lattice NRQCD simulations.  This assumes
that perturbative corrections at momentum scales beyond the lattice cutoff
(i.e. the inverse lattice spacing) are small.  Nonperturbative contributions
to the $c_i$ coefficients are not neglected, and appear explicitly in 
Eqs.~(\ref{action}-\ref{dH3}) via the tadpole factors, mentioned below.

A tilde on any quantity indicates that the
leading classical discretization errors have been removed.  Besides this
classical improvement, a mean-field form of quantum improvement is also
implemented through the introduction of two ``tadpole factors'', $u_s$ and
$u_t$, following the well-known suggestion of Lepage and Mackenzie\cite{LepM}.
Notice that the action of Eqs.~(\ref{action}-\ref{dH3}) allows different 
lattice spacings, $a_s$ and $a_t$, in the spatial and temporal directions;
hence two tadpole factors are required.

Various choices for the light quark and gauge field action, $S_{q,G}$ of 
Eq.~(\ref{action}), have been considered.  
In the lattice NRQCD simulations of Refs.~\cite{ourSwaves,ourPwaves}, the 
gauge field portion is classically and tadpole-improved,
whereas the authors of Refs.~\cite{AliKhan,Hein} chose the unimproved
Wilson gauge field action.  For the fermion portion, 
Refs.~\cite{ourSwaves,AliKhan,Hein} employ the tadpole-improved clover 
action\cite{SW} while Ref.~\cite{ourPwaves} uses a more highly improved
D234 action\cite{D234}.  When the heavy quark mass is sufficiently large
and the lattice spacing sufficiently small (but not small compared to the 
inverse heavy quark mass),
one expects the predictions of these actions to agree.

\section{S-WAVE HEAVY-LIGHT MASSES}

\begin{figure}[thb]
\epsfxsize=400pt \epsfbox[80 400 980 730]{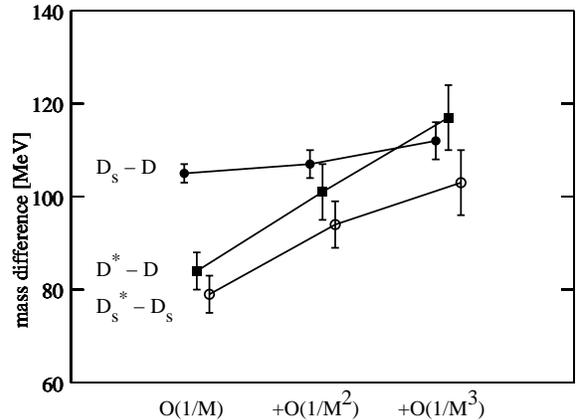}
\caption{S-wave splittings of charmed mesons at first, second and third order
         in NRQCD from Ref.~\protect\cite{ourPwaves}.  
         Solid lines simply connect related data points; they are not fits.}
\label{fig:sdiff}
\end{figure}
Lattice NRQCD simulations typically produce a clear signal for the
S-wave heavy-light meson masses, so they are convenient observables to use
for testing the heavy quark expansion.
Since the leading contribution to the heavy quark mass does not appear
explicitly in the NRQCD action, only mass differences (or more generally
energy differences) can be computed.  The absolute mass scale can be
extracted from the kinetic energy of a meson,
\begin{equation}
E_{\bf p}-E_0=\frac{{\bf p}^2}{2M_{\rm kin}}-\frac{{\bf p}^4}{8M_{\rm kin}^3}
+\ldots,
\end{equation}
to the desired order in $1/M_{kin}$, where $E_{\bf p}$ and $E_0$ are taken
from lattice simulations, and the ``kinetic'' mass $M_{kin}$ is thus obtained.

The first three orders in the $1/M$ expansion of 
Eqs.~(\ref{action}-\ref{dH3}) have been studied in 
Refs.~\cite{ourSwaves,ourPwaves} for two different light quark actions.
The charmed results of Ref.~\cite{ourPwaves} are displayed in 
Fig.~\ref{fig:sdiff}.  In this plot, subleading corrections to the $D_s-D$ 
splitting are insignificant, and the physical result agrees
well with the experimental result\cite{PDG},
\begin{equation}
\left[D_s-D^+\right]_{\rm expt} = 104 {\rm ~MeV}.
\end{equation}

The $D^*-D$ splitting of Fig.~\ref{fig:sdiff} has $O(1/M^2)$ and $O(1/M^3)$ 
corrections which are roughly equal, at about 20\% each, and the $D_s^*-D_s$
splitting acquires 20\% and 10\% corrections at $O(1/M^2)$ and $O(1/M^3)$
respectively.  Firm statements about convergence, or lack thereof, for the
$1/M$ expansion are difficult to support with only these data.  The first
correction terms are not far from their expected size, 
$O(\Lambda_{\rm QCD}/M_{\rm charm})$, but the $O(1/M^3)$ corrections are
perhaps larger than might have been hoped.
These spin splittings are clearly smaller than the experimental 
values\cite{PDG}
\begin{eqnarray}
\left[D^{+*}-D^+\right]_{\rm expt} &=& 141 {\rm ~MeV}, \\
\left[D_s^*-D_s\right]_{\rm expt} &=& 144 {\rm ~MeV},
\end{eqnarray}
as typical of quenched lattice 
calculations.\cite{ourSwaves,Boyle,ourPwaves,AliKhan}

The $1/M$ expansion
is nicely suited to the bottom mesons and, even with 2000 gauge field
configurations, the simulations of Ref.~\cite{ourPwaves} found the nonleading
contributions to be smaller than statistical uncertainties.
The quenched results are shown beside the experimental data in 
Table \ref{tab:B}.
\begin{table}
\caption{Quenched lattice NRQCD results for the S-wave bottom 
         mesons\protect\cite{ourPwaves} compared to experimental 
         data\protect\cite{PDG}, in units of MeV.
         }\label{tab:B}
\begin{tabular}{ccc}
        & quenched & experiment \\
\hline
$B_s-B$ & 92$\pm$3 & 90$\pm$2 \\
$B^*-B$ & 25$\pm$2 & 46 \\
$B_s-B$ & 26$\pm$1 & 47$\pm$3 \\
\hline
\end{tabular}
\end{table}

\section{P-WAVE HEAVY-LIGHT MASSES}

It is more difficult to extract a signal for the P-wave masses from lattice 
simulations.  Creation operators which extend over more than a single lattice
site need to be tuned to enhance the signal.

\begin{figure}[htb]
\epsfxsize=400pt \epsfbox[80 70 980 750]{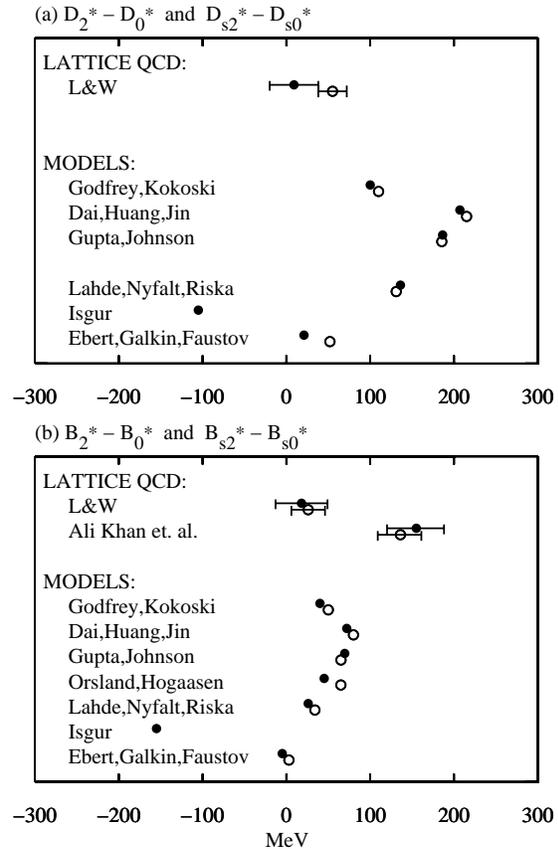}
\caption{Splittings between the $P_0$ and $P_2$ masses for the $D^{**}$
         and $B^{**}$ systems.  Open(solid) symbols involve an s(u,d) quark.}
\label{fig:models}
\end{figure}
In Ref.~\cite{ourPwaves}, the P-wave masses were studied at $O(1/M)$,
$O(1/M^2)$ and $O(1/M^3)$ for both charmed and bottom mesons.  Subleading
contributions were found to be smaller than the statistical uncertainties.
Fig.~\ref{fig:models} compares the splitting between two P-wave masses (J=0
and J=2) as obtained from lattice calculations and from various 
models.\cite{models}
For the $B^{**}$ system, most methods predict a splitting between zero and
80 MeV.

The discrepancy between the two lattice results is somewhat disconcerting.
Both computations employed quenched lattice NRQCD and both had a temporal
lattice spacing of about 0.1 fm.  Ref.~\cite{ourPwaves} used an improved
gauge action with a spatial lattice spacing of 0.2 fm, a highly-improved 
(D234) light-quark action and an ensemble of 2000 configurations.  
Ref.~\cite{AliKhan} used the unimproved (Wilson) gauge action with a spatial
lattice spacing near 0.1 fm, the improved clover action for light quarks and
an ensemble of 102 configurations.

\begin{figure}[thb]
\epsfxsize=400pt \epsfbox[80 70 980 750]{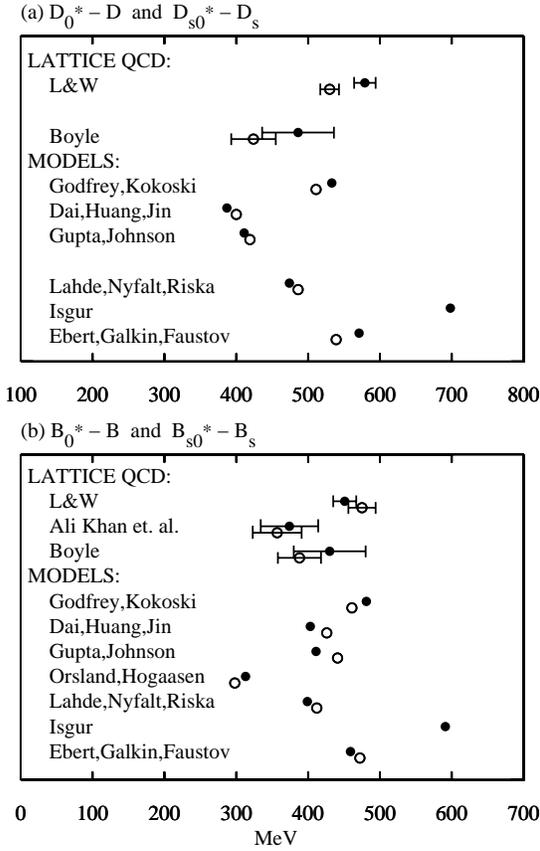}
\caption{S-P splittings for heavy-light mesons.  
         Open(solid) symbols involve an s(u,d) quark.}
\label{fig:modelP0}
\end{figure}
In Ref.~\cite{ourPwaves}, the ratio of $P_2/P_0$ correlation functions
falls monotonically as Euclidean time increases, and of course statistical
uncertainties grow as Euclidean time increases.  Thus, if the meson mass is
extracted from a region which begins too near the initial time, the predicted
mass splitting would be too large.
With this in mind, it is noted in Ref.~\cite{ourPwaves} that, even at 
the smallest Euclidean times,
those lattice data cannot reproduce the large splitting of Ref.~\cite{AliKhan}.
Instead,
Ref.~\cite{ourPwaves} arrives at a plateau-independent upper bound of 100 MeV
for the $B_2^*-B_0^*$ splitting.
Lattice efforts are ongoing, and we look forward to a clearer understanding
of the $D^{**}$ and $B^{**}$ spectra.

In Fig.~\ref{fig:modelP0}, the splitting between S and P-wave mesons is
shown.  The lattice results of Boyle are interesting in that they do not rely
on NRQCD.\cite{Boyle}

\section{SUMMARY}

The convergence of lattice NRQCD has been studied up to $O(1/M^3)$ for both
S and P-wave heavy-light mesons.  The only cause for concern seems to be
the S-wave spin splitting for charmed mesons, where a definitive statement
cannot yet be made about convergence.

The P-wave masses have recently come under study with lattice NRQCD.
The results of Ref.~\cite{ourPwaves} indicate that the $D^{**}$ splittings
are $O(50 {\rm ~MeV})$ or smaller, while the $B^{**}$ splittings
are $O(10 {\rm ~MeV})$ or smaller.  However, the lattice work reported in 
Ref.~\cite{AliKhan} is not consistent with this, so further studies will
be instructive.

This work was supported in part by the Natural Sciences and Engineering
Research Council of Canada.

\end{document}